\documentclass[preprint,12pt,authoryear]{elsarticle}
\usepackage{amssymb}
\usepackage{amsmath}
\usepackage{mathtools}
\usepackage{bm}
\usepackage{xcolor}
\usepackage{url}

\usepackage{credits}

\usepackage{graphicx}
\usepackage{hyperref}
\hypersetup{
     colorlinks=true,
     linkcolor=blue,
     filecolor=blue,
     citecolor = orange,      
   }
\usepackage[dvipsnames]{xcolor}
\usepackage{siunitx}

\journal{International Journal of Multiphase Flow}
\begin{document}
\begin{frontmatter}

\title{Inferring the Turbulent Breakup of Colloidal Aggregates \\Using Graph Neural Networks}

\author[label1,label2]{M. Buzzicotti}
\affiliation[label1]{
organization={Dept. Physics, University of Rome “Tor Vergata”},
addressline={Via della Ricerca Scientifica 1},
city={Rome},
postcode={00133},
country={Italy}
}
\affiliation[label2]{
organization={INFN, Sez. Roma 2},
addressline={Via della Ricerca Scientifica 1},
city={Rome},
postcode={00133},
country={Italy}
}
\author[label3,label2]{M. Cencini}
\affiliation[label3]{
organization={Istituto dei Sistemi Complessi, CNR ISC},
addressline={Via dei Taurini 19},
city={Rome},
postcode={00185},
country={Italy}
}
\author[label1,label2]{G. Cimini}

\author[label4]{M. Vanni} 
\affiliation[label4]{
organization={Dip. Scienza Applicata e Tecnologia DISAT, Politecnico di Torino},
addressline={Corso Duca degli Abruzzi, 24},
city={Torino},
postcode={10129},
country={Italy}
}

\author[label5,label6]{A. S. Lanotte\corref{cor1}}
\ead{alessandrasabina.lanotte@cnr.it}
\affiliation[label5]{
organization={Institute of Nanotechnology, CNR NANOTEC},
addressline={Via Monteroni},
city={Lecce},
postcode={73100},
country={Italy}
}
\affiliation[label6]{
organization={INFN, Sez. Lecce},
addressline={Via Arnesano},
city={Lecce},
postcode={73100},
country={Italy}
}
\cortext[cor1]{Corresponding author}
\date{}

\begin{abstract}
Solid aggregates in turbulent suspensions may break under the action
of shear stresses. We explore the use of Graph Neural Networks (GNN)
to infer aggregate fragmentation once the aggregate structure and flow
velocity gradients are known. We consider two models: the first GNN is
a classifier, trained to distinguish aggregates that break from those
that do not; the second GNN is a regression model, trained to predict
the maximal tensile force within each aggregate in a given flow
condition. We show that both models complete their task with a high
statistical accuracy, also generalizing to aggregates of different
sizes, and generally performing better than the statistical prediction
based on mean field quantities. This work paves the way for future use
of GNN to quantify aggregate breakup in a large population of
aggregates suspended in complex flow configurations, as it takes place
in the wet production of fine powders and in the transport of
sediments in environmental flows.

\end{abstract}
\end{frontmatter}

\section{Introduction}
In turbulent suspensions, the breakup of agglomerates -- material
structures formed by an ensemble of fragments or primary particles or
monomers compacted together -- due to the hydrodynamic stress
generated by the surrounding fluid is a key process in a wide range of
applications, in both natural and engineering contexts. Examples
include rubber compounding \citep{Manas2012}, the production of
micro-particles via wet synthesis \citep{Gavi2010}, the rheological
behavior of suspensions \citep{Bilodeau1998}, and sediment transport
in natural flows or wastewater streams \citep{Kaveh_2025}. Despite the
broad scope of these applications, only a few experimental studies
have analyzed the breakup process at the scale of individual
agglomerates \citep{Harshe_2011,Saha_2015,Saha_2016,Coletti2025}. Most
experimental work focuses instead on the global properties of the
population of particles -- such as the particle size distribution or
mean size --, rather than on the mechanism and evolution of individual
breakup events, the reason being that agglomerates breakup is very
complex and strongly depends on the aggregate structure and flow-field
conditions over time. As a result, such studies do not provide tools
for accurate prediction of agglomerate breakup processes.

This limitation can be addressed through numerical simulations that
couple fluid dynamics with contact mechanics to model aggregate
behavior. Several approaches have been proposed in the literature for
this purpose. Many of these (based, e.g., on Discrete Element Methods
(DEM) and/or Immersed Boundary methods) for the turbulent breakup of
aggregates demand complex implementation and considerable
computational resources \citep{Saxena2025,Chen2020,Leonelli2026}. As a
consequence, their application soon becomes impractical for large
populations of agglomerates when directly coupled with complex flow
fields obtained by integrating the three-dimensional Navier-Stokes
equations. On the modeling side, there have been attempts to derive
scaling laws for the critical stress required to initiate breakup in
sheared or turbulent colloidal suspensions (see e.g.,
\cite{Zaccone2009}).

An alternative strategy, which we explore here, is to develop
surrogate models based on neural networks (NN), capable of determining
the breakup condition of an agglomerate after appropriate
training. These models are generally computationally efficient and
could enable simulations involving very large and heterogeneous
aggregate populations, thereby providing a realistic representation of
process equipment, thus representing a promising avenue of numerical
modeling. Recently, \citet{Khalifa2023} trained an artificial NN to
quantify collision-induced breakage and agglomeration processes of
initially nearly spherical agglomerates, and where the DEM simulations
were carried out in vacuum. The trained model was then integrated into
an Eulerian–Lagrangian simulation framework to study collision-induced
breakup in T-junction laminar and turbulent flows, replacing the
physical collision model with the efficient neural surrogate
\citep{Khalifa2024}.

In this study, we consider turbulent dilute suspensions, focusing on
the role of shear stresses, which we solve exactly (within grid
resolution) and that possess a wide distribution with large tails in
homogeneous flow conditions \citep{Babler2015}. We present a
data-driven approach, based on Graph Neural Networks (GNN), which is
meant to make predictions on aggregate breakup without performing
simulations to directly compute the forces acting on the
aggregate. GNN represent a particularly suitable learning paradigm for
predicting aggregate fragmentation, because a colloidal aggregate can
be naturally framed as a graph-like object: primary particles interact
only through the contact bonds established during aggregation, and the
transmission of hydrodynamic stresses throughout the structure depends
on this connectivity.

In the last few years, GNN
\citep{GNNmodel,hamilton2017inductive,7974879} have become a standard
tool for machine learning on graph-structured data, that is, data
represented as a set of entities (nodes) with features, and of a set
of relationships (links) among them. Initially mainly used to solve
node classification and link inference tasks \citep{ZHOU202057},
nowadays GNNs have a wide span of applications across domains,
including chemistry and material science
\citep{Reiser2022,Thomas2023,Corso2024}. In fluid dynamics, GNNs have
been used for different goals. Since they operate on graphs, which can
be converted from any mesh geometry, GNNs can ensure a significantly
better resolution for relevant regions within the domain with the same
mesh size (see e.g. \citet{Barwey_2025}), compared with standard
convolutional neural networks. Interestingly, the graph structure
allows to study particles dynamics in Stokes suspensions, enabling the
construction of surrogate models for the mobility tensor of particles
subject to hydrodynamical interactions and external forces
\citep{Ma2022}. While GNNs have been used to simulate the complex
fluid dynamical problems under a wide range of conditions
\citep{pmlr-v119-sanchez-gonzalez20a,Se2024,Gao2024,Filiatraut23,Serhani2024},
including GNN-based LES schemes \citep{Serhani2024} or multiphase
flows in porous media \citep{Vu2025}, their application to solid
  aggregates in turbulent suspensions remains a relatively unexplored
  research area.

Here we develop two different GNN models: the first one is simply
meant to classify whether a single aggregate will break or not; the
second is meant to infer the maximal tensile force acting on one of
its internal links. Using the latter model, the inferred maximal force
can then be compared to a threshold value for rupture to establish
whether the aggregate will break or not.  We deal with small
agglomerates, whose breakup is induced by the hydrodynamic stresses
generated by the turbulent flow only
\citep{kusters1997aggregation,debona2014}, rather than by the
combination of different processes (flow stresses, inter-agglomerate
collisions or impacts with the walls). The GNN is
  trained with the results of a physical model of the process based on
  a joint Direct Numerical Simulation and Stokesian dynamics – DEM
  approach that examines rigid isostatic agglomerates with a fixed
  fractal dimension of immersed in 3D homogeneous and isotropic
  turbulence. We mostly focus on a systematic investigation of
  aggregates of given size (i.e. with a fixed number of monomers, see
  below), albeit with different geometries. This choice is linked to
  the objective of the work, which is to demonstrate the applicability
  of GNNs to the prediction of aggregate breakup in a
  three-dimensional turbulent suspension, a complex system
  characterized by both large-scale coherent structures and highly
  non-Gaussian and intermittent small-scale fluctuations. Then, we
  extend our analysis to an heterogeneous population of aggregates, to
  test the generalisation ability of our model. Extension to turbulent
  suspensions in confined volumes will be evaluated in future work
  based on the results obtained here.

The paper is organized as follows. In Sec.~\ref{sec:flow}, we
introduce the system under consideration and the numerical modeling of
the turbulent suspension. Then, in Sec.~\ref{sec:stokesian}, we
details the databases (DBs) used to train and test the GNNs. The Graph
Neural Network models - classifier and regression- are introduced in
Sec.~\ref{sec:models}, while a statistical test to be compared to the
Neural Networks results is presented in Sec.~\ref{sec:test}. Section
\ref{sec:results} is devoted to discussing the performances of the two
GNNs on different datasets for homogenenous in size suspensions, the
origin of model failures and the generalisation ability when
considering heterogeneous in size suspensions. We summarise our work
in Section \ref{sec:final}.

\section{Numerical modeling of turbulent suspensions}
\label{sec:flow}
The simplest numerical methods to study aggregate breakup use the
free-draining approximation, where each primary particle is assumed to
be unaffected by the flow perturbations caused by neighboring
particles
\citep{Becker_2009,Eggersdorfer_2010,Schutte_2018,Ruan_2020}. Estimates
of the screening effect of hydrodynamic forces within agglomerates
have been based on the fraction of the primary particle surface area
directly exposed to the external fluid flow
\citep{Higashitani_2001,Fanelli_2006}. More sophisticated
semi-empirical methods, that relate screening to the solid volume
fraction or aggregate geometry, were later adopted by \citet{Yao_2021}
and \citet{Yu_2023}. These approaches fall within the
Eulerian–Lagrangian framework, in which aggregates are treated as
point-like entities.

A more computationally demanding alternative involves full resolution
of the fluid flow at the scale of the primary particles. This route
was taken by different authors by means of Lattice-Boltzmann
simulations to investigate the restructuring and breakup of
agglomerates in shear \citep{Saxena2022,Saxena2023}, accelerated
\citep{Saxena2025} and turbulent flows \citep{Derksen_2013}.\\ For
colloidal aggregates with very small particle Reynolds number,
$Re_p=\gamma_{eff}R^2/\nu \ll 1$ ($R$ being the aggregate size in
terms of its radius of gyration, $\nu$ the fluid kinematic viscosity
and $\gamma_{eff}$ the effective shear rate), Stokesian Dynamics (SD)
offers an equally accurate alternative, enabling precise computation
of the hydrodynamic forces acting on the aggregates. Compared to
techniques that fully resolve the solid–liquid interfaces, and the
hydrodynamic forces/torques acting on each primary particle, SD is
less computationally expensive and allows for the study of systems
with several hundreds of primary particles, in contrast to full CFD
simulations, which typically handle only a few dozen of them. Over the
years, Stokesian dynamics has been used to investigate the breakup and
restructuring of soft and rigid agglomerates in uniform shear flows
\citep{Harada_2006,Seto_2011,VanniGastaldi2011,Harsche_2012,Frungieri_2021},
elongational flows \citep{Ren_2015}, complex confined flows
\citep{Vasquez_2022,Frungieri_2022}, and isotropic homogeneous
turbulence \citep{debona2014}. For aggregates that remain rigid,
without deforming, until the onset of breakup, the Stokesian dynamics
simulation approach can be significantly accelerated using the
procedure proposed by \citet{Vanni2015}. \\

We consider a colloidal suspension of small aggregates whose center of mass, ${\bf X}_{cm}(t)$, evolves as neutrally buoyant tracer particle 
\begin{equation}
    \frac{d{\bf X}_{cm}}{dt}= {\bf V}_{cm}(t)\,,
    \label{eq:tracer}
\end{equation}
${\bf V}_{cm}(t)={\bf v}({\bf X}_{cm}(t),t)$ being the fluid velocity at
the aggregate center of mass. Equation \ref{eq:tracer} is valid for
aggregates of size $R$ much smaller than the Kolmogorov scale of the
turbulent flow, $\eta$, and whose density matches that of the
fluid. Moreover, we consider a very dilute suspension, where aggregate
collisions, hydrodynamic coupling and particle feedback onto the flow
can be neglegted \citep{Balachandar_2010}.

The fluid phase evolves according to the three-dimensional incompressible (${\bf \nabla} \cdot {\bf v}=0$) Navier-Stokes equations,
\begin{eqnarray}
\partial_t {\bf v} + ({\bf v}\cdot {\bf \nabla}) {\bf v} &=& - \frac{1}{\rho_f}{\bf \nabla} p + \nu \nabla^2 {\bf v} + {\bf f}\,,
\label{eq:NS}
\end{eqnarray}
where $p$ is the pressure field, $\rho_f$ is the fluid density and
${\bf f}$ is an external three-dimensional forcing. At statistical
steady state, the power input, $\epsilon_{in}= \langle {\bf f} \cdot
{\bf v}\rangle$, balances kinetic energy dissipation, $\epsilon_{\nu}=
\nu\langle |{\bf \nabla}{\bf v}|^2\rangle$, where angular brackets
stand for spatial average over the fluid volume.

Direct Numerical Simulations (DNS) of the Navier-Stokes equations are
solved adopting a standard pseudo-spectral approach, fully dealiased
with the two-thirds rule, within a cubic, periodic domain of length $2
\pi$. The kinematic viscosity is such that $\eta \simeq \Delta x$,
i.e. the Kolmogorov length-scale $\eta$ is of the order of the grid
spacing $\Delta x$. This implies that turbulent dissipative range
dynamics is well resolved. The statistically homogeneous and isotropic
random forcing $f({\bf x},t)$, continuously supplying kinetic energy
to the flow, acts at large scales, around the wave-number $k_f=1$, and
it is the solution of a second-order Ornstein–Uhlenbeck process
\citep{sawford1991reynolds}. In our numerical set-up,
  there is no imposed mean shear, and the effective shear rate is
  estimated as $\gamma_{eff}= \sqrt{\epsilon_{\nu}/\nu}$ (see later
  on). Relevant parameters of the DNS are summarised in Table
  \ref{table:dns} and are given in simulation units. \\
\begin{table}
    \centering
    \begin{tabular}{cccccccccccc}
        \hline
        $Re_{\lambda}$ & $u_{rms}$ & $\varepsilon_{in}$ & $\gamma_{eff}$ &$\nu$ & $\eta$ & $L$ & $T_E$ & $\tau_{\eta}$ & $dt_{Lag}$ & $T_{tot}$ & $N_p$ \\
        $90$ & $3.3$ & $2.3$ & $15.1$ & $0.01$ & $0.024$ & $\pi$ & $0.95$ & $0.07$ & $0.006$ & $36$ & $524000$\\
        \hline
    \end{tabular}
    \caption{Parameters of DNS in simulation units (su). Microscale
      Reynolds number $Re_{\lambda}$ corresponding to a resolution
      $N^3=256^3$ grid points; root-mean-square velocity $u_{rms}$;
      the mean kinetic energy input rate $\epsilon_{in}$ equal to the
      mean kinetic energy dissipation rate $\epsilon_{\nu}$;
      root-mean-square velocity gradient $\gamma_{eff}=
      \sqrt{\epsilon_{\nu}/\nu}$; kinematic viscosity $\nu$;
      Kolmogorov length-scale $\eta = (\nu^3/\epsilon_{\nu})^{1/4}$;
      integral scale $L$, Eulerian large-scale eddy turnover time $T_E
      = L/u_{rms}$, Kolmogorov timescale $\tau_{\eta}=
      \sqrt{\nu/\epsilon_{\nu}}$; sampling time along Lagrangian
      trajectories $dt_{Lag}$,; total integration time $T_{tot}$,
      total number of advected Lagrangian tracers $N_p$.}
    \label{table:dns}
\end{table}
Aggregates moving in the turbulent flow are subject to internal
stresses due to the interaction of the aggregates with the local flow
field. In the colloidal suspension here considered, since $Re_p\ll 1$,
significant variations in the fluid velocity gradients take place on
distances much larger than the size $R$ of the aggregates: this means
that the particle size is assumed to be much smaller than the
Kolmogorov scale of the flow, $R\ll \eta$. Consequently, in
evaluating the hydrodynamic stresses, it is reasonable to neglect the
curvature of the velocity profiles and assume that the aggregates are
surrounded by a linear flow field with uniform velocity
gradient. Under this assumption, Stokesian dynamics \citep{Brady1988}
can be effectively adopted to numerically determine the hydrodynamic
forces, torques and stresses acting on each aggregate. The method is
mesh-less and provides the relationship between the hydrodynamic
interactions acting on the particles and their velocity. The velocity
$\mathbf{V}_i$ at each of the i-$th$ primary particle position
$\mathbf{x}_i$ is given by
\begin{equation}
\label{eq:aggr_vel}
\mathbf{V}_i = \mathbf{V}_{cm}^{\infty} + \mathbf{\Gamma}^{\infty} \cdot (\mathbf{x}_i - \mathbf{X}_{\mathrm{cm}}),
\end{equation}
where $\mathbf{V}_{cm}^{\infty}$ is the undisturbed flow velocity and
$\mathbf{\Gamma}^{\infty}$ is the velocity gradient tensor,
$\Gamma_{ij}= \partial_j v_i$, evaluated at the aggregate center of
mass. Additionally, the aggregates are assumed to be rigid, and
hydrodynamic forces -- associated to both translational and rotational
motions -- are redistributed as internal stresses over the aggregate
\citep{VanniGastaldi2011,Vanni2015}. SD simulations allow to evaluate
the stress distribution and the location where the stress is highest,
i.e. the bond for the onset of breakup.\\ The results presented here
are obtained by first solving the aggregates motion as point-like
tracers evolving in the three-dimensional turbulent flow, and
computing the instantaneous value of the velocity gradient matrix at
their position. The SD dynamics is then integrated offline to compute
the maximal tensile force. Further details can be found in
\citep{debona2014}.\\
\begin{figure}
\begin{center}
\includegraphics[clip=true,keepaspectratio,width=0.5\textwidth]{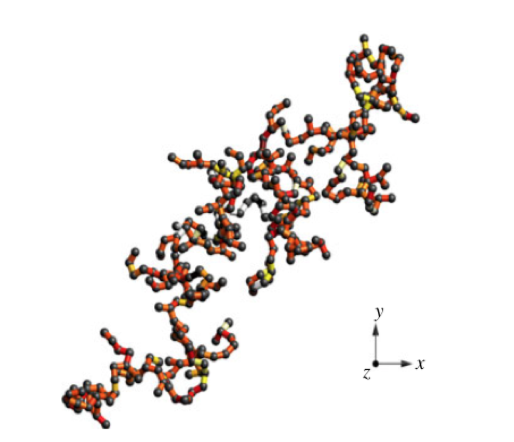}  
\end{center}
\caption{\label{fig:aggregate} Structure of a cluster-cluster
  aggregate of fractal dimension $D_F=1.9$, made up on $N_M=384$
  primary particles of radius $a$, and $N_L=N_M-1$ links among
  them. The gyration radius of the aggregate scales with the primary
  particle radius and the number of primary particles as $R \simeq a
  \,N_M^{1/D_F}$.}
\end{figure}
We consider colloidal suspensions of rigid and isostatic aggregates
made up by few hundreds of primary particles of radius $a$. They are
generated numerically by using a tunable cluster-cluster method
\citep{Lattuada2003}, capable of producing aggregates which do not
have spherical symmetry and obey the following relation
\begin{equation}
\label{eq:CC}
N_M \simeq \left( \frac{R}{a}\right)^{D_F}\, 
\end{equation}
where $N_M$ is the number of primary particles composing the
aggregate, and $R$ is the gyration radius of the
aggregate. For a systematic study, we set $D_F=1.9$
  and $N_M=384$, and the typical shape is given in
  Fig.~\ref{fig:aggregate}, then we will briefly consider aggregates
  with different values of $N_M$ at the end of our work. Agglomerates
  with such features are quite common, as they are typically obtained
  by fast coagulation processes of sub-micron primary particles. The
  value of fractal dimension is a measure of the mechanism of
  aggregation, with smaller values being associated with more open
  structures. The condition of isostaticity implies that there are no
  closed loops of particles in the structure and all the chains of
  primary particles in the aggregate are open on one side, so that the
  failure of a single bond results in the breakup of the aggregate
  into two fragments.

\section{Aggregates statistics and ground-truth data}\label{sec:stokesian}
When aggregates move and rotate in the turbulent flow, the
hydrodynamics stresses produce a turnover between compression and
elongation forces acting on each inter-particle link. For isostatic
aggregates, breakup occurs whenever the maximal internal tensile force
(i.e. on the most loaded bond) exceeds the critical pull-off value,
given by contact mechanics \citep{Johnson1985}, $F_{cr}\propto a
\sigma$, where $\sigma$ is the surface energy at the contact. Hence,
to determine the occurrence of breakup the key observable is the
maximal force $F_{max}$ to be compared to the pull-off threshold
value. Since breakup occurs upon the failure of a single bond,
aggregates with multiple bonds with local tensile force above the
critical threshold are considered nonphysical.\\
\begin{figure}
\begin{center}
\includegraphics[width=1.\textwidth]{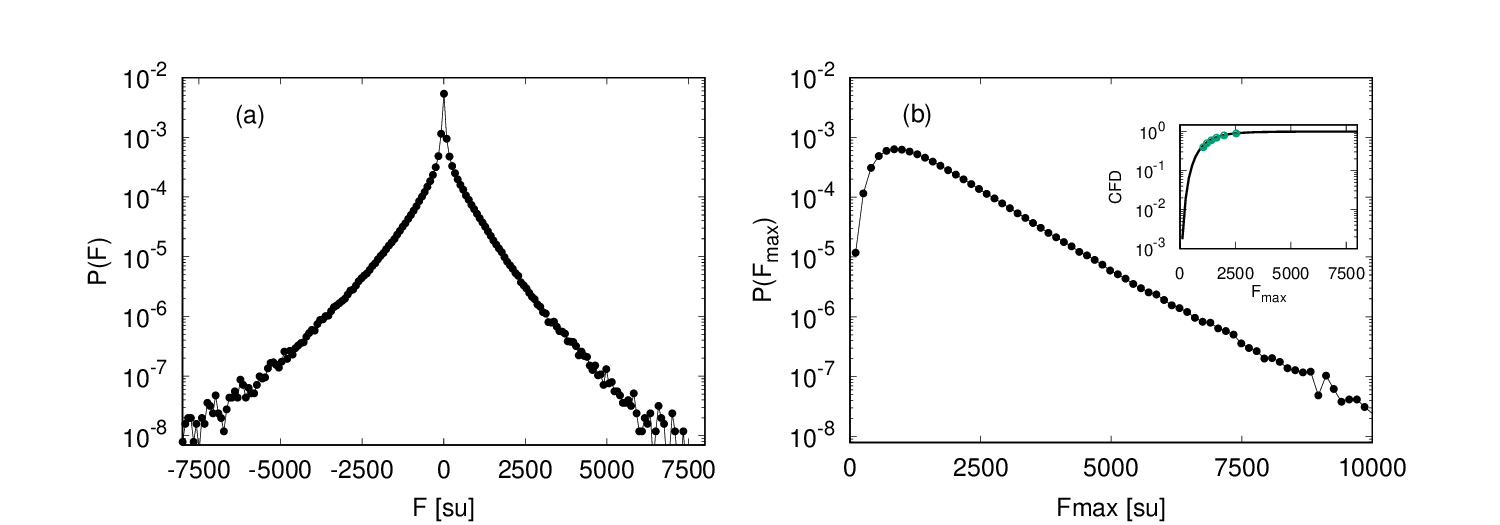} 
\end{center}
\caption{\label{fig:PDF_CDF} (a) PDF of the tensile force $F$,
  exhibiting tails with exponential behaviour \citep{debona2014}. (b)
  PDF of the maximal tensile force $F_{\text{max}}$; in the inset, the
  cumulative distribution of $F_{\text{max}}$(continuous line). The
  green circles plotted on top of the CDF identify the six threshold
  values for the maximal tensile force, $F_{th1} = 1052$;
  $F_{\text{th2}} = 1220$; $F_{\text{th3}} = 1417$; $F_{\text{th4}} =
  1653$; $F_{\text{th5}} = 1988$; $F_{\text{th6}} = 2539$,
  corresponding to the CDF being equal to 0.4, 0.5, 0.6, 0.7, 0.8 and
  0.9, respectively.}
\end{figure}
In Fig.~\ref{fig:PDF_CDF}(a), we plot the probability density function
(PDF) of the normal tensile force averaged over all the aggregates in
the suspension, that exhibits a stretched exponential behaviour and is
symmetrical. We note that at the level of the single bond, this PDF
may be skewed towards negative values (if compression prevails) or
positive values (if extension prevails).  Figure~\ref{fig:PDF_CDF}(b)
shows the PDF of the maximal tensile force, $F_{\text{max}}$, while
the inset displays its cumulative distribution function (CDF). From
the CDF, six threshold values for the breakup are defined,
corresponding to the CDF ranging, with equal spacing, from 40\% to
90\% (as detailed in the figure caption).  These thresholds define six
distinct aggregate families, such that for each of them breakup occurs
when the tensile force on the most loaded bond, $F_{\text{max}}$
exceeds the threshold value, i.e., $F_{\text{max}} > F_{\text{th}}$.

For our purpose of developing a data-driven tool, it is important to
construct unbiased datasets, containing random, independent
realizations. This helps to avoid temporal correlations that could
possibly influence the NN learning stage. From DNS data, we selected
$2 \times 10^6$ velocity gradient realizations, that are statistically
independent. Additionally, we considered 1000 different aggregate
shapes, all characterized by the same fractal dimension $D_F$ and
number of monomers $N_M$, but differing in shapes. Each
aggregate-shape was then randomly rotated through 2000 different
orientations, and each configuration was exposed to a different
velocity gradient. Finally, for each of these realizations, we
computed the Stokesian dynamics, and obtained the tensile force on
each aggregate bond from which we extract the value for the most
loaded bond, $F_{\text{max}}$. These data have been further split into
different datasets for the learning and testing stages, summarised in
Table \ref{tab:1}. Details of the dataset preparation can be found in
Appendix A. \\
\begin{figure}[ht!]
\centering
\includegraphics[width=0.6\textwidth]{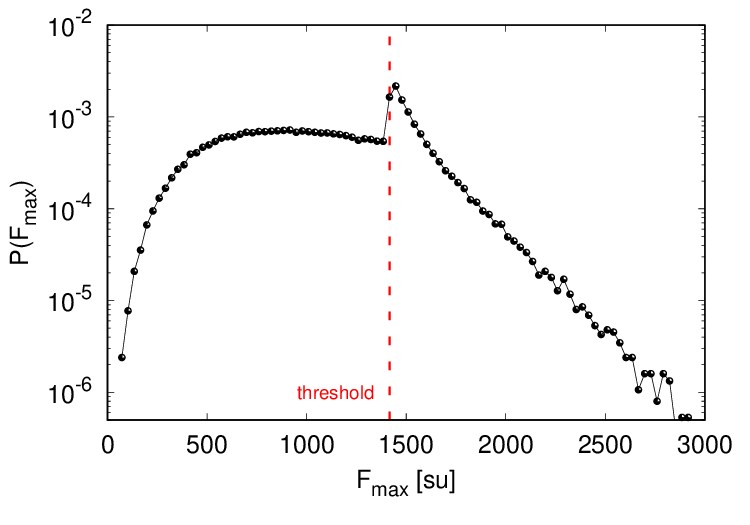}
\caption{\label{fig:PDF_wthreshold} PDF of the maximal tensile force
  $F_{\text{max}}$ for the aggregates that either do not break, or
  break for a single bond failing when $F_{\text{max}} \ge
  F_{\text{threshold}}$, with $F_{\text{th3}}=1417$ (vertical red
  dashed line).}
\end{figure}

\subsection{Database description: Classification task}
\label{subsec:DB_CL}
To classify broken and unbroken aggregates, we selected realizations
for which $F_{\text{max}}$ exceeds a threshold value on a single bond
only, in order to respect the isostatic assumption of the aggregates:
by doing so, we generated a database containing 720.000 examples out
of which 576.000 where used as learning dataset. These are further
organized into six families, corresponding to the different rupture
thresholds. In other words, each (single-rupture) aggregate has been
selected by measuring the maximal tensile force over all the internal
bonds and comparing it to one randomly selected threshold value from
the six selected values.

As explained above, after having generated a large DB of $2 \times
10^6$ aggregates, we impose the threshold for the rupture and
eliminate all aggregates that possess more than one bond with maximal
tensile force above threshold. This procedure is physically justified
by the fact in a physical experiment, as soon as the maximal force
reaches the threshold value the aggregate breaks and that it is very
unlikely that two bonds overcome the threshold value at the same
instant. As a consequence, as we do not follow the evolution of
aggregates but generate them from random rotations and uncorrelated
instances of the velocity gradients, those samples that present more
than one rupture should be considered as not physically realizable, as
they are expected to be already broken. From a statistical point of
view this procedure causes a conditioning that modifies the statistics
of the maximal tensile force (Fig.~\ref{fig:PDF_CDF}(b)). This effect
is exemplified in Fig.~\ref{fig:PDF_wthreshold}. Most of the broken
aggregates experience a value of $F_{\text{max}}$ close to the
threshold value, and the resulting PDF decreases very sharply as the
value of $F_{\text{max}}$ increases above the threshold value. Hence,
for broken aggregates, the maximal tensile force is mostly
concentrated just above the pull-off value for the bond
failure. Differently, the distribution of the unbroken aggregates is
broader, since for these aggregates any force smaller than the
threshold ensures the absence of a link failure. However, the
probability close to the threshold is modified by the threshold value
also for the unbroken aggregates and this impacts the performance of
the GNN classifier, as we will see later in the result section. This
is generally true for all thresholds, and it is related to the
single-bond rupture mechanism.

To mitigate the statistical prevalence of broken cases very close to
the threshold (which are the most challenging to predict), and reduce
the statistical imbalance of broken versus unbroken cases close the
threshold values, we opted for the following choice: the learning
database contains 33\% occurrences of broken aggregates and 67\%
occurrences of unbroken aggregates for each of the six
thresholds. After the learning or training stage, the GNN classifier
performances were tested on three different DBs, that we briefly
describe in the following (see also Table \ref{tab:1}).

{\bf Cl-imb:} A test-set was generated using velocity gradients and
random rotation of the aggregates that were never seen during
training. This test-set is unbalanced, consistently with the training
DB, and contains approximately 67\% unbroken aggregates and 33\%
broken aggregates.

{\bf Cl-bal:} A balanced test-set containing 50\% broken, 50\%
unbroken aggregates. It is obtained by randomly selecting a subset of
realizations of the previous dataset, Cl-imb.

{\bf Cl-rnd:} A balanced test-set (50\% broken, 50\% unbroken), made
of realizations exhibiting a single rupture with respect to a randomly
selected threshold value for the maximal tensile force. In particular,
for each sample, a random threshold value is drawn with uniform
probability in the range [5\%; 95\%] of the CDF of the maximal tensile
force (corresponding to the interval $F_{\text{max}} \in [432;
  3100]$). After this random selection, the threshold is compared to
the maximal tensile force computed by Stokesian Dynamics, with the
constraint that no more than a link has tensile force above threshold,
to determine if it is a broken or unbroken aggregate realization. The
puropose of this DB is to test the classifier model ability to
generalize with respect to threshold values never seen during the
training.\\

\subsection{Databases description: Regression task}
\label{subsec:DB_Re}
To train and test the GNN regression model aimed to infer the maximal
tensile force for each aggregate in a given flow realization, we
prepared another dataset. In the regression model, the rupture
threshold is not known and there is no \textit{a priori} distinction
between broken and unbroken aggregates. The learning database contains
525.000 samples obtained from a collection of about one thousand
different aggregate shapes, exposed to different rotations and fluid
velocity gradients; the GNN regression model was then tested on the
following two DBs.

{\bf Re-test:} A test-set with about 132.000 samples of aggregates
exposed to velocity gradient and rotation realizations, never seen
during the training. The geometrical shapes of the aggregates
composing this test database have been seen by the model during the
training stage.

{\bf Re-gen:} A test-set with about 128.000 samples of aggregates
whose shape, velocity gradient and rotation realizations had never
been seen during the training stage. Hence, this dataset provides a
basis for evaluating the extent to which the GNN can generalize its
performance to unseen geometries of the aggregates. The new shapes
still have the same fractal dimension, $D_F$, and the same number of
primary particles, $N_M$, as the aggregates used for training. \\

\section{Graph Neural Network models for aggregate breakup}
\label{sec:models}
\begin{figure}[ht!]
\centering
\includegraphics[width=1.\textwidth]{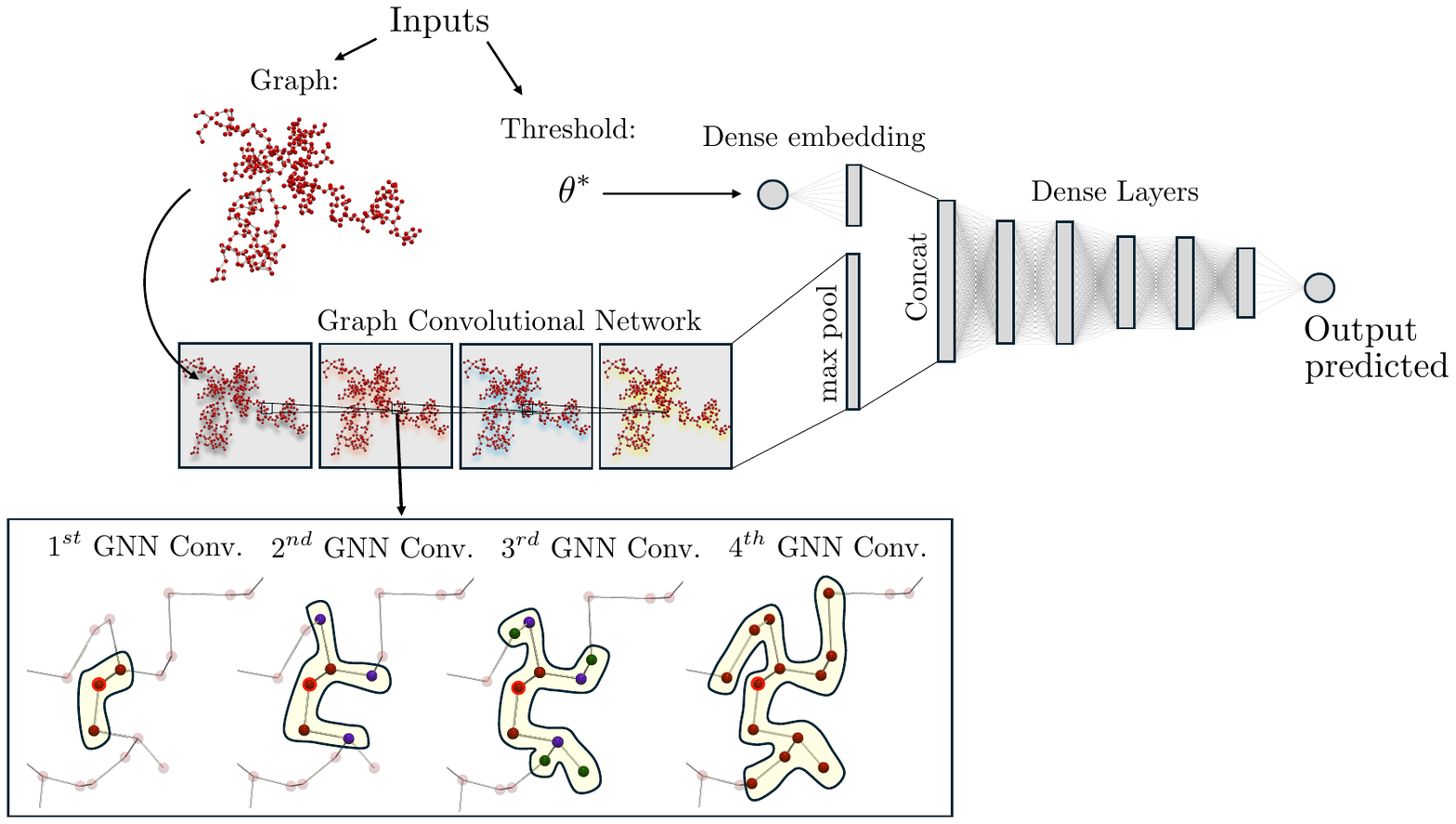}
\caption{\label{fig:PDF_GNNscheme} A scheme of the end-to-end
  prediction task with the GNN classifier producing a binary output;
  for the regression model, the scheme is very similar, but there is
  no threshold value given in input, and the output is a positive real
  number.}
\end{figure}

The basic architecture of the two models, namely the {\it classifier}
and the {\it regression} GNNs, is sketched in
Fig.~\ref{fig:PDF_GNNscheme}: the models share several general
features and most architectural implementation details. In particular,
the input data consist of an aggregate represented as a graph, where
each monomer corresponds to a node and the connections are encoded in
a bidirectional adjacency matrix, whose elements
denote physical contact between particles. For each $i$-th node
$i= 1, \dots, N_M$, the input features form a six-dimensional vector
\begin{equation}
  (x_i, y_i, z_i, V_{x,i}, V_{y,i}, V_{z,i})\,,
\label{eq:features}
\end{equation}
with the node’s position coordinates relative to the aggregate’s
center of mass $\mathbf{X}_{\mathrm{cm}}$, and the velocity at each
primary particle position. The velocity at each node is given by
Eq.~\eqref{eq:aggr_vel}.

Differently from the regression model, the classifier additionally
receives the threshold value $F_{\mathrm{th}}$ corresponding to
pull-off force. The threshold information is encoded through a fully
connected layer that takes the one-dimensional real value as input and
maps it into a 64-dimensional latent vector. This information is
essential for predicting the rupture, whereas it is not required for
estimating the maximum tensile force experienced by the aggregate's
links. In the latter case, the likelihood of breakup can be evaluated
\textit{a posteriori} by comparing the inferred maximal tensile force
with the threshold value.

In the training stage, the classifier learns to distinguish between
broken and unbroken aggregates with respect to six different threshold
values of the maximal tensile force, as discussed above.
The input graph, in both models, is processed through four successive
concatenations of graph convolutional layers. Each convolution is
performed according to the GraphSAGE
scheme~\citep{hamilton2017inductive}, with a max aggregation operation
applied after each convolution. Specifically, the node embeddings are
updated at the $m$-th layer as
\begin{equation}
\label{eq:embedding}
\mathbf{h}_i^{(m)} = \mathbf{W}_1^{(m)} \cdot\mathbf{h}_i^{(m-1)} + \mathbf{W}_2^{(m)} \cdot \text{max} \left\{ \mathbf{h}_j^{(m-1)} : j \in \mathcal{N}(i) \right\},
\end{equation}
where $\mathbf{h}_i^{(m)}$ denotes the embedding of node $i$ at layer
$m$, $\mathcal{N}(i)$ is the set of neighbors of node $i$ (the
particles connected to $I$ by cohesive bonds), $\text{max}$ indicates
element-wise maximum aggregation. $\mathbf{W}_1^{(m)},
\mathbf{W}_2^{(m)}$ are two learnable weight matrices that project the
6-dimensional input feature vector (\ref{eq:features}) into a
64-dimensional embedding space. In our framework, we do not apply a
non-linear activation function after each layer.\\ At the end of the
four graph convolutions, the output at each node is constructed by
concatenating the four 64-dimensional feature vectors obtained from
each layer, resulting in a final 256-dimensional representation for
each node. At the final stage of the GNN convolutions, a global max
pooling operation is applied over all node embeddings to obtain a
single 256-dimensional latent vector representing the entire input
graph. This pooling operation is independent of the number of nodes in
the input graph enabling, in principle, the architecture to process
aggregates of varying sizes.\\
\begin{figure}[ht!]
\centering
\includegraphics[width=1\textwidth]{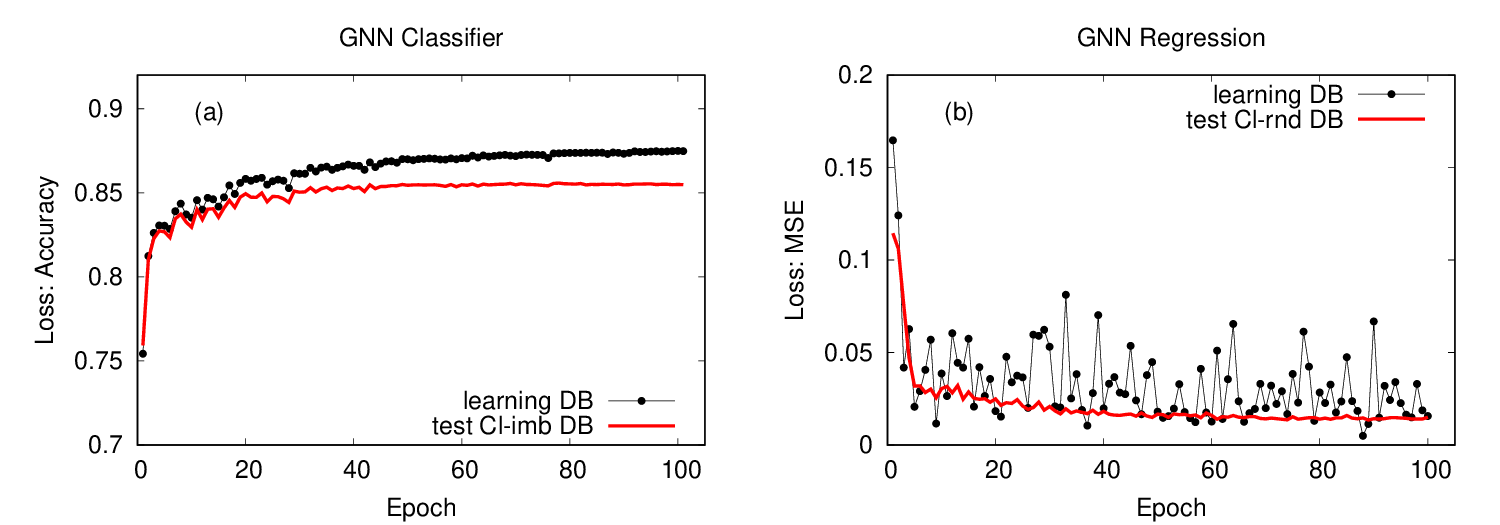}
\caption{\label{fig:loss} (a) The evolution of success probability (or
  accuracy) with the number of epochs for the GNN classifier model on
  the learning dataset (filled circles) and the test Cl-imb dataset
  (continuous red line). (b) The evolution of loss function,
  $L_{MSE}$, with the number of epochs for the GNN regression model on
  the learning dataset (filled circles) and the test Cl-rnd dataset
  (continuous purple line).}
\end{figure}
In the classifier, the 64-dimensional embedding, encoding the
threshold information, is concatenated at this point with the
256-dimensional output of the graph embedding, resulting in a combined
320-dimensional latent vector.  At this stage, the input embeddings
for both models are mapped onto the desired one-dimensional output
through six fully connected layers. These layers gradually reduce the
input dimension while applying a nonlinear activation function. The
first five dense layers use a ReLU activation, while the activation of
the final output layer is model-specific, in particular: the
classifier employs a softmax activation, while the regression model
uses a linear activation.\\

The {\it classifier} GNN outputs a rupture probability in $[0,1]$,
with $1$ denoting breakage. Training consists in minimising the binary
cross-entropy loss, $L_{CE}$, between the model rupture probability
and the ground-truth label $Y^{true}$\,:
\begin{equation}
    L_{CE}(p, Y^{true}) = - \frac{1}{N}\sum_{k=1}^{N} [Y_k^{true} \log(p_k)-(1 - Y_k^{true}) \log(1 - p_k)]\,,
    \label{eq:binary_cross_entropy}
\end{equation}
where $N$ is the batch size in the learning process.\\ The {\it
  regression model} is trained to predict the maximal tensile force
exerted by the fluid on the aggregate, outputting a positive real
number. In this case, during the training the normalised mean squared
error (MSE) loss, $L_{MSE}$, is minimised\,:
\begin{equation}
    L_{MSE}=\frac{1/N\sum_{k=1}^N [(F^{pred}_k - F^{true}_k)^2]}{\langle (F^{true})^2\rangle_{DB}} =\frac{\langle (F^{pred} - F^{true})^2\rangle }{\langle (F^{true})^2\rangle_{DB}}\,, \label{eq:MSE1}
\end{equation}    
where $N$ is the batch size in the learning process, and the average
$\langle \bullet\rangle_{DB}$ is over the whole learning
database.\\ Figure~\ref{fig:loss} illustrates the evolution of
accuracy for the classifier (left) and mean squared error,
eq.(\ref{eq:MSE1}), for the regression model (right), over training
epochs for the training and test datasets. The curves obtained during
GNN testing still vary across epochs because test loss and accuracy
are repeatedly computed during training after each update to the
network’s parameters. In this workflow, the model is optimized
exclusively with respect to the loss evaluated on the training data,
while the test datasets are used to periodically assess performance.
Monitoring both the training and test metrics is essential to verify
that the learning process is converging. A decrease in training loss
alone is indeed insufficient, because the model may improve on the
training set due to overfitting, resulting in degraded
generalization. By comparing the two training and testing curves, one
can identify when the model stabilizes and determine if the
performance on unseen data reflects the improvements observed during
training.

\section{Statistical test for aggregate break-up}\label{sec:test}
In order to benchmark the performances of the GNN models for the
classification of aggregate breakup, we devised a simple statistical
model for detecting rupture, which is illustrated in the following.

In our set-up, the breakup of aggregates is due to hydrodynamic
stresses acting on the links between the primary particles composing
the aggregate \citep{kusters1997aggregation,debona2014}. In simple
shear flows, the hydrodynamic stress is given by the uniform shear
rate, $\gamma$. In turbulent flows, it can be quantified in terms of
the effective shear rate
\begin{equation}
\label{eq:shear}
\gamma_{eff} ({\bf x},t) \equiv \sqrt{ S_{ij}S_{ij}}\,, 
\end{equation}  
where summation over $i,j$ is implied and $S_{ij}$ denotes the
symmetric part of the velocity gradient tensor, i.e.
$S_{ij}=(\Gamma_{ij}+\Gamma_{ji})$ and $\Gamma_{ij}=\partial_j
v_i$. As it is well known, we can write
$\gamma_{eff}=\sqrt{\epsilon_{\nu}/\nu}$, where $\epsilon_{\nu}({\bf
  x},t)$ is the local kinetic energy dissipation that, in turbulence,
is known to exhibit strong fluctuations both in space and time
\citep{frisch1995turbulence,meneveau1991multifractal,benzi2009velocity}. \\ For
brittle isostatic aggregates, as considered here, we assume that
breakup takes place as soon as $\gamma_{eff}$ is larger than a given
threshold $\gamma_{cr}$. Actually, it has been shown that turbulent
breakup may not only depend on the instantaneous value of
$\gamma_{eff}$, but also on the instantaneous orientation of the
aggregate with respect to the flow field \citep{debona2014}. For
simplicity, we ignore such an effect.

\begin{figure}
\centering
\includegraphics[clip=true,keepaspectratio,width=0.6\textwidth]{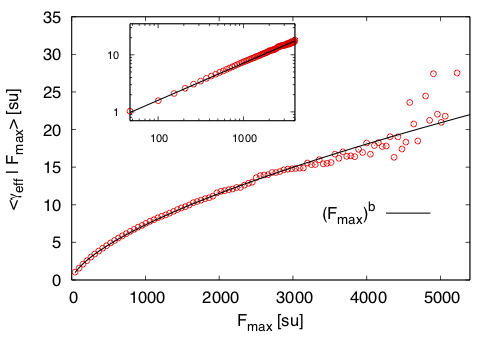}
\caption{\label{fig:epsfmax} Conditional effective shear rate $\langle
  \gamma_{eff}|F_{\text{max}}\rangle$ vs $F_{\text{max}}$ computed on
  the learning dataset of the GNN classifier ($576.000$ samples). The
  solid curve shows the fitted behavior,
  eq.(\ref{eq:empirical}). In the inset, the same
    plot in log-log scale, where a linear fit is applied in the range
    $F_{max}\in [20:3000]$. The fit gives the exponent $b=0.65 \pm
    0.02$ and the prefactor $c=0.08\pm 0.01$.}
\end{figure}

We computed the effective shear rate $\gamma_{eff}$ conditioned on the
value of $F_{\text{max}}$, $\langle \gamma_{eff}|F_{max}\rangle$, in
the same dataset used to train the GNN-Cl model. The result is plotted
in Fig.~\ref{fig:epsfmax}. We observe that such a quantity behaves in
very good approximation as a power law,
\begin{equation}
\langle \gamma_{eff}|F_{\text{max}}\rangle\propto c\, F_{\text{max}}^{b}\,,
\label{eq:empirical}
\end{equation}  
with the exponent $b \approx 2/3$, and $c$ a prefactor.  We note that
a dimensional relation between the tensile force and the effective
shear rate is expected for sub-Kolmogorov aggregates, namely $F
\propto \mu R_g^2 \gamma_{eff}$, where $\mu$ is the dynamic viscosity
of the flow. However, this dimensional argument does not constrain how
the maximal force value depends on the effective shear rate.\\ While
further investigation is needed to better assess the observed power
law, we recall that relationships of this kind are known in turbulent
aggregates, e.g. the typical size of an aggregate is known to be
proportional to an inverse power of the average effective shear
strength (see, e.g., the review of \citet{jarvis2005review} and
references therein).\\
The observed power law of eq.~(\ref{eq:empirical}) can be used to
establish a statistical test for the breakup of an aggregate, provided
one knows, $\gamma_{eff}$, and the threshold force,
$F_{\text{th}}$. These inputs are the same that the GNN classification
model receives, since the effective shear rate $\gamma_{eff}$ is
implicitly available through the flow velocities at the positions of
the primary particles. In particular, it is natural to assume that the
aggregate breaks when $\gamma_{eff} > c F_{\text{th}}^{b}$, while it
remains unbroken otherwise.  Applying the above statistical test on
the test-set Cl-imb, we find that this test works correctly for
$\approx 74.46\%$ of the samples, while in Cl-rnd $69.69\%$. The
difference between the performances in the two DB may be due to the
fact that the statistical test and in particular the constant $A$ in
eq.~(\ref{eq:empirical}) was fitted on the data with 6 thresholds and
so there may be some tiny statistical differences from the random
threshold DB.

\section{Results and discussion}
\label{sec:results}
In this section we will first present the results of the two models in
classifying broken and unbroken aggregates in the test databases, then
we will discuss the origin of the errors in classification specific of
the two models and finally we will details the performances of the
regressor model in predicting the maximal tensile force.

\subsection{Classification task} 
\label{subsec:classify}
As discussed in the previous section, the classifier model GNN-Cl
directly outputs the rupture probability. In particular, we remark
that the GNN-Cl makes a correct prediction when the output is either
1.AND.$F_{\text{max}} \ge F_{\text{th}}$ or 0.AND.$F_{\text{max}} <
F_{\text{th}}$. In order to use the GNN regression model for
classifying the aggregates, we proceed in two steps. First, the
regression model outputs the value of the maximal tensile force, for
each sample. Then, the predicted maximal force is compared with the
assigned threshold for the specific realization to establish whether
the predicted value is above (broken) or below (unbroken) threshold.
The discussion on the regression ability, i.e. the performances of the
GNN-Re in predicting the correct $F_{\text{max}}$, will be postponed
to the next subsection, here we focus on the classification task.
\begin{table}[!h]
\centering
\resizebox{\columnwidth}{!}{%
\begin{tabular}{|c|c|c|c|c|c|c|c|c|c|c|c|}
\hline
DB & Model & Loss & Tot & Broken & Unbroken & Correct & Wrong & BCU & UCB & \%success & F$_1$-score\\ 
\hline        
Cl-imb  &  Cl & CE & 144000 & 47789 & 96211 & 123143 & 20857 &  8043 &  12814  & 85.5 & 0.79 \\
Cl-imb  & Re & MSE & 144000 & 47789 & 96211 & 115295 & 28705 &  21925 &  6780  & 80.0 & 0.64\\
Cl-imb  & SM & -- & 144000 & 47789 & 96211 & 107218 & 36647 &  23257 &  13390  & 74.5 & 0.57 \\
\hline 
\hline
Cl-bal &   Cl & CE & 90000 & 45000 & 45000 & 76454 & 13546  & 7562  & 5984   & 84.9 & 0.84\\
Cl-bal &   Re & MSE& 90000 & 45000 & 45000 & 66282 & 23718 & 20621  &  3097  & 73.6 & 0.67\\
\hline 
\hline
Cl-rnd & Cl & CE& 200000 & 100000 & 100000 & 166604 & 33396  & 22441 & 10955 & 83.3 & 0.82\\
Cl-rnd & Re & MSE & 200000 & 100000 & 100000 & 150082 & 49918  & 44879 &  5039 & 75.0 & 0.68\\
Cl-rnd & SM & -- & 200000 & 100000 & 100000 & 139371 & 60629  & 48986 &11643 & 69.7 & 0.62 \\
\hline 
\end{tabular}
}
\caption{Col 1: name of the test database; col 2: Model name: Cl and
  Re correspond to the GNN classifier and regression models,
  respectively, while SM corresponds to the statistical model
  discussed in Sec.~\ref{sec:test}; col 3: Loss function based on
  binary cross-entropy (CE) or mean-squared error (MSE); col 4: total
  number of aggregate samples; col 5 : number of broken aggregates;
  col 6 number of unbroken aggregates; col 7 : total number of correct
  predictions; col 8: total number of wrong predictions; col 9: false
  negatives, or BCU, i.e. true Broken aggregates Classified as
  Unbroken; col 10: false positives or UCB, true Unbroken Classified
  as Broken; col 11: (unconditional) success
    probability, giving the correct prediction over the total number
    of samples; col 12: F$_1$-score, giving the harmonic mean of
    precision and recall. Note the different composition of the two
  databases: Cl-imb contains about $2/3$ of unbroken aggregates and
  $1/3$ of broken ones; differently, Cl-bal and Cl-rnd are both
  equally split.
 \label{tab:1}}
\end{table}

Table~\ref{tab:1} summarizes all results, together with information
about the models, datasets and loss functions used during the training
and test stages. In particular, the success probability is defined as
the number of correctly classified samples normalised to the total
number of samples. We see that the GNN-Cl works very well, correctly
classifying at best 85.5\% of the samples, while failing in the 14.5\%
of the cases. Also in all datasets, the GNN-Cl works better than the
GNN-Re, which at best correctly classifies 80\% of the
samples. Finally, we can see that the statistical test of
Sec.~\ref{sec:test} has poorer results compared to both the GNN
models.\\ Besides the success probability,
  Table~\ref{tab:1} presents the so-called F$_1$-score or F-measure,
  giving the predictive performances for the classification of broken
  aggregates \citep{fawcett2006introduction}. The F$_1$-score is the
  harmonic mean of precision and recall,
  \begin{equation}
  F1=2 \, \frac{p\times r}{p+r}
  \label{eq:F1}
  \end{equation}
  where the precision $(p)$ is the ratio between the number of
  aggregates correctly classified as broken (true positive) and all
  aggregate that were predicted to be broken (all positive), and the
  recall $(r)$ is the ratio between the number of aggregates correctly
  classified as broken (true positive) and those which are actually
  broken (true positive plus false negative). A high value of the
  F$_1$ score means that the classifier is both precise and sensitive:
  results show that the GNN-Cl outperforms the classification results
  of both the regression and the statistical models. Moreover, we
  observe that for both the DB CL-imb and CL-bal, the success
  probability and the F$_1$ values remain high. This confirms that
  having trained the GNN-Cl with an unbalanced dataset does not impact
  model performances: when applied to balanced test dataset, the
  GNN-Cl still performs very well.\\
  
\begin{figure}[ht!]
\centering
\includegraphics[width=1.\textwidth]{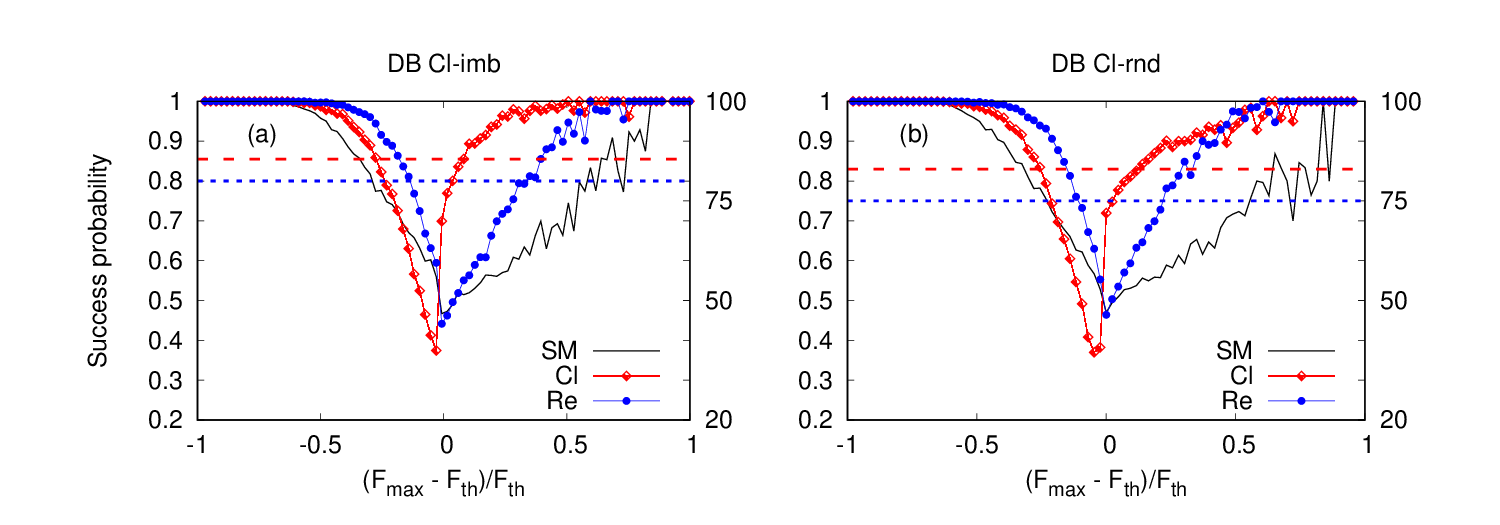}
\caption{\label{fig:1} Success probability (in classification) as a
  function of the signed relative difference,
  $(F_{\text{max}}-F_{\text{th}})/F_{\text{th}}$, between the maximal
  tensile force $F_{\text{max}}$ and the threshold value,
  $F_{\text{th}}$. (a) Models are tested on dataset Cl-imb; (b) the
  same but for dataset Cl-rnd. In each plot, the red curve with
  diamonds refers to the GNN Classifier (Cl), the blue curve with
  filled circles refers to the GNN Regression (Re) model, and the
  continuous line refers to the statistical model. The color
  horizontal dashed lines correspond to the global success probability
  on each model (col 11 in Table \ref{tab:1}), colors are the same of
  the model lines.}
\end{figure}

To disentangle instances where the GNNs fail, we look at wrong results
as follows: we define {\it false positive} the unbroken aggregates
that the GNN classifies as broken (UCB), and {\it false negatives}
broken aggregates that the GNN classifies as unbroken (BCU). These
data are also reported in Table~\ref{tab:1}.  We can see that in
general the regressor model has a larger probability to detect false
negatives, while the classifier is more balanced in Cl-imb and Cl-bal
while it also produces more false negative in the DB Cl-rnd. Also the
statistical test tends to generate more false negative than
positive.\\As shown in Fig.~\ref{fig:PDF_wthreshold}, once a threshold
is assigned and the condition to have at maximum one bond with the
tensile force above threshold is imposed, the PDF of the maximal force
strongly deviates from the unconditional PDF, shown in
Fig.~\ref{fig:PDF_CDF}(a). In particular, the form of the PDF is
strongly modified close to the threshold and for values above it. It
is thus natural to expect that the distance of the maximal tensile
force from the threshold plays an important role in the classification
task. To better inspect this, we can look at the success probability
conditioned on the relative distance between the maximal tensile force
of each aggregate and the threshold force associated to the breakup
condition, $X=\frac{F_{\text{max}}-F_{\text{th}}}{F_{\text{th}}}$. In
Fig.~\ref{fig:1}, we show the conditional success probability of the
two GNN models and for the statistical model, for the datasets Cl-imb
and Cl-rnd in the left and right panel, respectively. For each model,
datapoints that lie above the dataset average success rate (dashed
color lines) refer to samples for which the model predictions work
better than its average.  As expected, correct predictions are more
and more frequent as the distance from the threshold value is larger
and larger: in other words, when the true force is much larger (or
smaller) then the assigned threshold, the model correctly predicts the
rupture (or its absence, respectively).\\ The region of force values
close to the threshold is the one where making correct predictions is
the most difficult, and all models concentrate their failures, but not
in the same way.

\begin{figure}[ht!]
\centering
\includegraphics[width=0.7\textwidth]{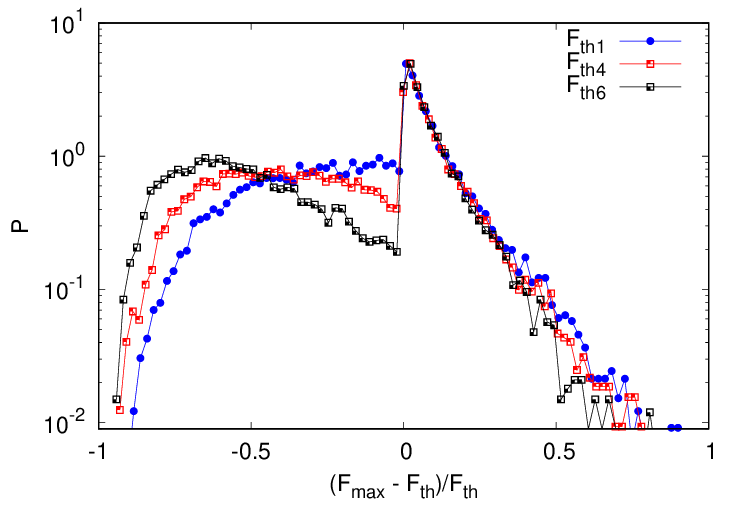}
\caption{\label{fig:probXsoglia} PDF of the relative deviation between
  the maximal force and the threshold, conditioned to the different
  threshold values, $F_{th1}, F_{th4}$ and $F_{th6}$. Data refer to
the GNN-Cl tested on the database Cl-bal.}
\end{figure}

First, we note that GNN-Cl is better than GNN-Re at classifying in
particular in the region above threshold, i.e. for broken
aggregates. In this region, the statistical test is the one performing
the worst. The GNN-Re model, as detailed in the next subsection, tends
to underestimate $F_{\text{max}}$ when its true value is large, and
this is the main source of false negatives.  However, we should take
into consideration two facts: first, regression models have been
trained and optimised to predict $F_{\text{max}}$ and not to classify;
second, regression models have been trained on an aggregate dataset in
which no threshold value was imposed, while classifier models have
been trained on samples such that there is only one link with tensile
force larger than a given threshold. This implies that the
distribution of the forces in the learning databases of the GNN-Cl and
GNN-Re models differ, and in particular the training dataset used for
the GNN-Re may contain aggregates with multiple links having a tensile
force above some given threshold.\\ In Fig.~\ref{fig:1}, we have seen
that the success probability conditioned on the relative distance from
the threshold is not symmetric. Indeed, for the GNN-classifier most of
the failures come from samples for which $F_{\text{max}}<
F_{\text{th}}$, i.e. unbroken aggregates, that have been wrongly
classified as broken. These cases, defined (UCB), are on the left part
of the distribution, and their success probability is much smaller
than the reference value.  To better understand the behavior of the
classifier we have studied separately the 6 families identified by
their thresholds.  In Fig.~\ref{fig:probXsoglia}, we show the
distribution of the relative distance between the true maximal force
and the threshold, $(F_{\text{max}}-F_{\text{th}})/F_{\text{th}}$, for
three of the six considered threshold values (the smallest, a central
one, the largest). We clearly observe that in the negative region,
i.e. when $F_{\text{max}}<F_{\text{th}}$, the smaller the threshold
the larger is the number of realizations with $F_{\text{max}} \approx
F_{\text{th}}$, while in the positive region (of broken aggregates)
the distribution is independent of the threshold value. This suggests
that the greater number of wrong predictions is expected for small
thresholds, while it should be less probable to misclassify, for
larger and larger thresholds.
In other words, the performances of the GNN classifier are hence
expected to depend on the threshold value, as it is confirmed by the
results for specific thresholds reported in Table~\ref{tab:DB1.1}. As
one can see for the largest threshold the success probability reaches
92\% while for the smaller one is only 78,3\% and
  similar discrepancies for the F$_1$-score.

We remark that the number of instances with larger threshold are less
than those with smaller threshold because of the condition of having
at maximum one bond above threshold, and the larger the threshold the
larger is the probability to have more than one bond above it.
\begin{table}[!h]
\centering
\resizebox{\columnwidth}{!}{
\begin{tabular}{|c|c|c|c|c|c|c|c|c|c|c|c|}
\hline
Th name & Th value & Tot  & Broken & Unbroken & Correct & Wrong & BCU & UCB & \%success & F$_1$-score\\ 
\hline
1 & 1052  & 15067  & 7618  & 7449  & 11803  & 3264 &  2055  & 1209  & 78.3 & 0.80\\
2 & 1220  & 14906  & 7362  & 7544  & 12125  & 2781 &  1599  & 1182  & 81.3 & 0.82\\
3 & 1417  & 14945  & 7486  & 7459  & 12437  & 2508 &  1389  & 1119  & 83.2 & 0.84\\
4 & 1653  & 15245  & 7540  & 7705  & 13081  & 2164 &  1120  & 1044  & 85.8 & 0.86\\
5 & 1988  & 14931  & 7516  & 7415  & 13287  & 1644 &  827   & 817   & 88.9 & 0.89\\
6 & 2539  & 14906  & 7478  & 7428  & 13721  & 1185 &  572   & 613   & 92.0 & 0.92\\
\hline
\end{tabular}
}
\caption{Performances of the classifier model GNN-Cl in the test-set
  Cl-bal, with results disaggregated by the values of the threshold on
  the maximal tensile force. Col 1: threshold name; col 2: value of
  the threshold; col 3: total number of aggregate samples; col 4 :
  number of broken aggregates; col 5 number of unbroken aggregates;
  col 6 : total number of correct predictions; col 7: total number of
  wrong predictions; col 8: false negatives, or BCU, i.e. true Broken
  aggregates Classified as Unbroken; col 9: false positives or UCB,
  true Unbroken Classified as Broken; col 10: (unconditional) success
  probability; col 11: F$_1$-score.\label{tab:DB1.1}}
\end{table}
Similar results are obtained in the test-set Cl-rnd, with random
thresholds. Here, however the presence of a larger number of wrong
predictions for small thresholds is further amplified: the smaller the
threshold the larger the probability to have $F_{\text{max}}\approx
F_{\text{th}}$ for unbroken aggregates. In other words, there are
proportionally more samples which are difficult to handle for the
GNN-Cl (see Fig.~\ref{fig:1}).\\ Summarising, we conclude that the
smaller the threshold the larger the number of wrong predictions by
the models, as confirmed by the F$_1$-score.\\

\subsection{Regression task}
\label{subsec:regress}

\begin{figure}[ht!]
\centering
\includegraphics[clip=true,keepaspectratio,width=1.\textwidth]{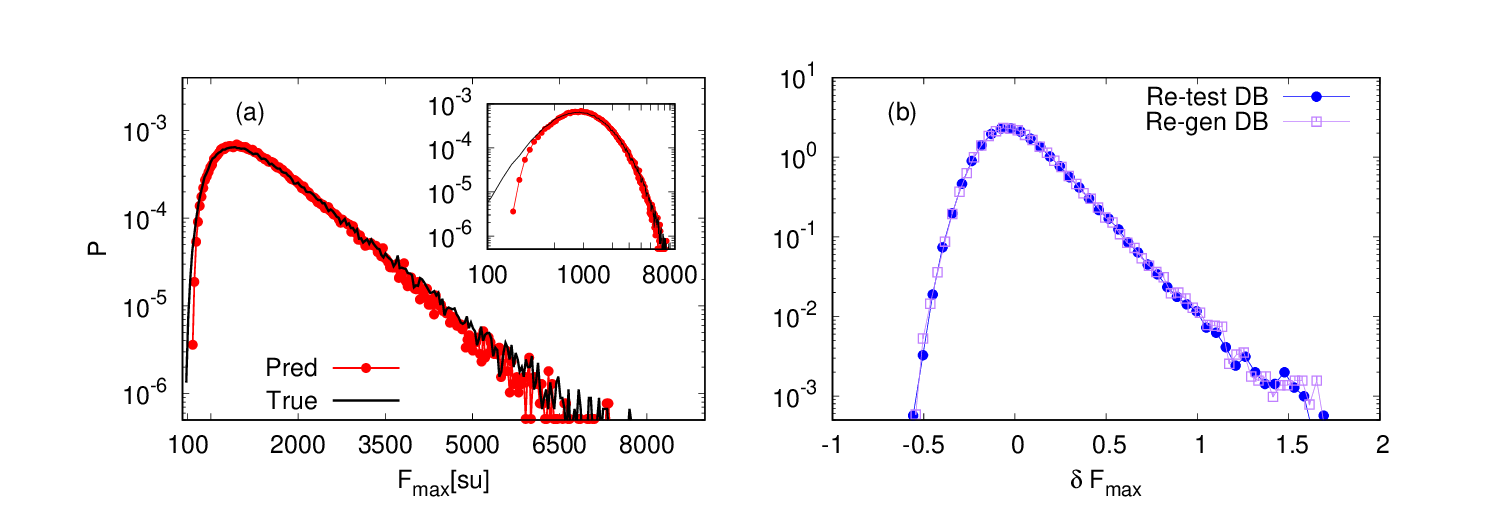}
\caption{\label{fig:fmax1} (a) PDF of the maximal tensile force (black curve), and of the one predicted by the GNN-Re (red circles), data refer to the test-set Re-gen; (inset) the same but in log-log coordinates, to better appreciate differences in the PDF tails. (b) the PDF of the relative error between the predicted maximal tensile force and the ground-truth one, $\delta F_\text{max}\equiv\frac{F_{\text{max}}^{\text{pred}}- F_{\text{max}}}{F_{\text{max}}}$, using the GNN-Re in two different test-sets: data from Re-test (blue circles) and from Re-gen (pink squares) (see definitions in Sec. \ref{subsec:DB_Re}).}
\end{figure}

We now discuss the performances of the GNN-Re model on the test-sets
Re-test and Re-gen (no threshold values, no selection of aggregates
with single bond rupture). Figure~\ref{fig:fmax1} (left) shows the
predicted distribution of the maximal tensile force against the
ground-truth one computed by the Stokesian dynamics in the test DB
made of the aggregates not seen during the learning. The predicted PDF
faithfully reproduces the true one, though some differences can be
appreciated in the tails.  In particular, we can see that in both the
right and left tails (see inset) the PDF of the predicted
$F_{\text{max}}$ is slightly below the ground truth meaning that
small/large $F_{\text{max}}$ tend to be over/underestimated.  This
effect is further quantified in Fig.~\ref{fig:fmax1} (right), where we
show distribution of the (signed) relative error of the predicted
maximal tensile force from the true one, ${\delta
  F_{\text{max}}}=\frac{F_{\text{max}}^{\text{pred}} -
  F_{\text{max}}}{F_{\text{max}}}$ for both the test-sets Re-test and
Re-gen. We observe that the distribution is not symmetric, i.e. for
meaning that the model rather overestimates and underestimate the
maximal tensile force.  Moreover, we also observe that there are no
appreciable differences between the results of the two test-sets,
suggesting a good generalization capability of the regression model.
\begin{figure}[ht!]
\centering
\includegraphics[clip=true,keepaspectratio,width=0.6\textwidth]{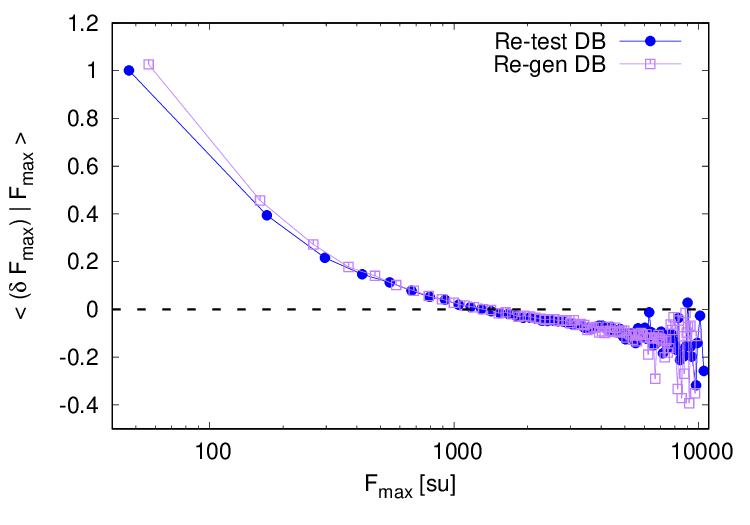}
\caption{\label{fig:condfmaxave} Mean relative error between the predicted maximal tensile force and the true one, obtained from Stokesian Dynamics, $\langle {\delta F_{\text{max}}}|\, F_{\text{max}} \rangle$. The GNN-Re model is tested on two different test-sets: data from Re-test (blue circles) and from Re-gen (purple squares).}
\end{figure}
To better disentangle the model behaviour we look at the mean (signed) relative deviation, $\langle {\delta F_{\text{max}}}|\, F_{\text{max}} \rangle$, conditioned on the value of the maximal tensile force.   
We clearly observe that the GNN-Re tends to overestimate the force when this is small and underestimate it when this is large; here, by {\it
  small} and {\it large}, we mean values of the maximal tensile force $F_{\text{max}}$ which are smaller/larger that the peak of the distribution (these are the values with zero the mean deviation). We also note that the zero crossing of the signed deviations (lower curves) is not an indication of a small error, but of the fact that for this value of the force the model can overestimate and underestimate the force with equal probability. Furthermore, one can see that on average the modulus of the relative error is larger for smaller $F_{\text{fmax}}$.

Finally, we compare the prediction by the GNN-Re model of the maximal tensile force distribution in the datasets built up with aggregates having the breakup of a single bond for a given threshold value, in the test-sets with fixed thresholds (Cl-imb) or random (Cl-rnd). As one can see the effect of the threshold on the PDF, signalled by the jumps in the PDFs, is smoothed out by the regressor model. Eventually, this can be cured by training the regressor on directly on the database used by the GNN-Cl and providing it the information on the threshold. \\
\begin{figure}[ht!]
\centering
\includegraphics[clip=true,keepaspectratio,width=1.\textwidth]{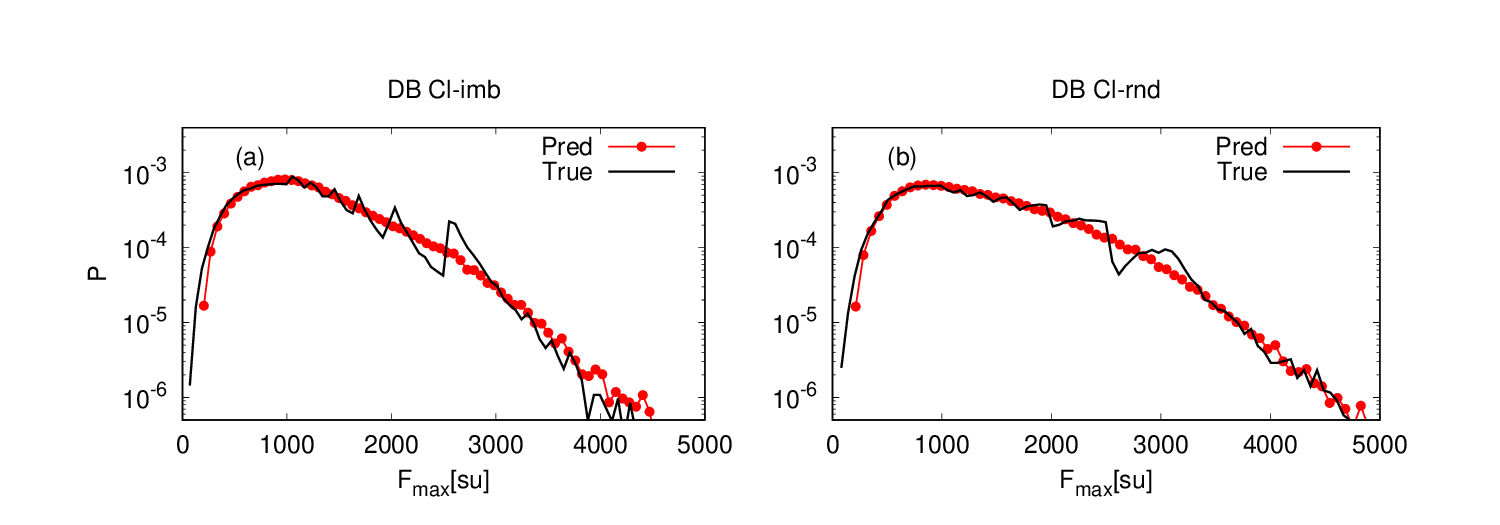}
\caption{\label{fig:DB1} (a) Comparison of the probability density
  functions of the ground-truth maximal force (continuous curve), and the one predicted by the GNN-Re in the test-set Cl-imb (purple squares);
  (b) the same but testing the GNN-Re on data of the DB Cl-rnd, with random thresholds (filled circles). The 'spikes' in the PDF of the ground-truth PDF correspond to the threshold values.}
\end{figure}

\subsection{Model heterogeneous Generalisation to different sizes}
\label{subsec:regress}
\begin{figure}
\begin{center}
\includegraphics[width=.6\textwidth]{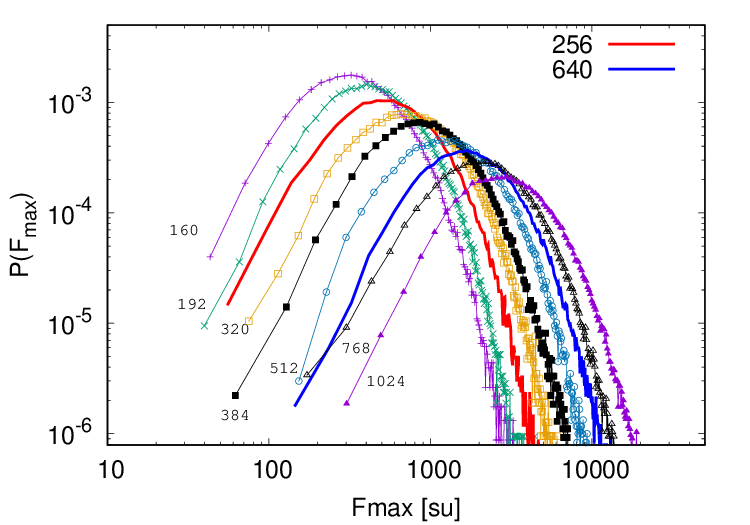} 
\end{center}
\caption{\label{fig:NM_variabile} PDF of the maximal tensile force
  $F_{\text{max}}$ for aggregates of different sizes (or
  families). Each family is defined by the number of primary particles
  composing the aggregates. Lines with symbols refer to families of
  the new training database $\mathrm{Cl_{NM}}$ (aggregates made with
  $N_M=160, 192, 320, 512, 768,1024)$, while continuous lines refer to
  families of the new heterogeneous testing database,
  $\mathrm{Cl_{heter}}$ (aggregates made with $N_M=256,640$).}
\end{figure}
Evaluating the ability of data-driven models to generalise to unseen
data is essential, as it indicates that they capture the underlying
relationships rather than simply memorizing the training data.  In the
previous section, we tested the models's capability to generalise to
unseen geometrical and velocity gradient realizations. Here we want to
push this further and consider the model ability to generalise to
aggregates of different sizes, namely aggregates with variable number
of primary particles. As it is clear from Figure
\ref{fig:NM_variabile}, aggregates of different sizes are
characterised by similar distributions of the maximal tensile forces,
but over quite different ranges of values.  This has an important
consequence: a GNN must be trained on a dataset containing aggregates
of heterogeneous sizes to learn the size-dependent tensile force
distributions; once trained, it can generalize to aggregate sizes not
encountered during training.  This is a very robust test of GNN
classifier ability to perform well in more complex suspensions.\\
\begin{table}[!h]
\centering
\resizebox{\columnwidth}{!}{%
\begin{tabular}{|c|c|c|c|c|c|c|c|c|c|c|}
\hline
DB & $N_M$ & Tot & Broken & Unbroken & Correct & Wrong & BCU & UCB & \%success & F$_1$-score\\ 
\hline        
$\mathrm{Cl_{NM}}$ & 160 & 12064 & 6028 & 6036 & 10453 & 1611 &  479 &  1132  & 86.6 & 0.87 \\
$\mathrm{Cl_{NM}}$ & 192 & 12043 & 5976 & 6067 & 10471 & 1572 &  400 &  1172  & 86.9 & 0.87 \\
$\mathrm{Cl_{NM}}$ & 320 & 11835 & 5939 & 5896 & 10219 & 1616 &  439 &  1177  & 86.3 & 0.87 \\
$\mathrm{Cl_{NM}}$ & 384 & 11937 & 5944 & 5933 & 10396 & 1541 &  470 &  1071  & 87.0 & 0.87 \\
$\mathrm{Cl_{NM}}$ & 512 & 12020 & 6018 & 6002 & 10646 & 1374 &  334 &  1040  & 88.5 & 0.89 \\
$\mathrm{Cl_{NM}}$ & 768 & 12019 & 5950 & 6069 & 10589 & 1430 &  418 &  1012  & 88.1 & 0.88 \\
$\mathrm{Cl_{NM}}$ & 1024 & 12082 & 5988 & 6094 & 10681 & 1401 &  423 &  978  & 88.4 & 0.88 \\
$\mathrm{Cl_{NM}}$ & all & 84000 & 41843 & 42157 & 73455 & 10545 &  2963 &  7582  & 87.4 & 0.88 \\
\hline
\hline
$\mathrm{Cl_{heter}}$ & 256 & 60000 & 30000 & 30000 & 52480 & 7520 &  1706 &  5814  & 87.4 & 0.88 \\
$\mathrm{Cl_{heter}}$ & 640 & 60000 & 30000 & 300000 & 52523 & 7477 &  2129 &  5384  & 87.5 & 0.88 \\
\hline 
\end{tabular}
}
\caption{Col 1: name of the test database; col 2: aggregate family
  type, defined in terms of the number of primary particles $N_M$
  within the aggregate; col 3: total number of aggregate samples; col
  4 : number of broken aggregates; col 5: number of unbroken
  aggregates; col 6: total number of correct predictions; col 7: total
  number of wrong predictions; col 8: false negatives, or BCU; col 9:
  false positives or UCB; col 10: (unconditional) success probability,
  giving the correct prediction over the total number of samples; col
  11: F$_1$-score, giving the harmonic mean of precision and
  recall. Note the different composition of the two databases:
  $\mathrm{Cl_{NM}}$ is a test dataset containing aggregate families
  seen during training; $\mathrm{Cl_{heter}}$ is a test dataset
  containing aggregate families never seen during training, and it is
  thus used to fully test generalisation ability of the
  GNN-Classifier.
 \label{table:NM}}
\end{table}
For this goal, we built new databases, for training and testing
respectively. For the GNN classifier model training, we consider a
well balanced dataset composed of aggregates with variable number of
primary particles $N_M = 160, 192, 320, 384, 512, 768, 1024$ with 40
different geometries for each $N_M$. \\ We collect 60000 samples for
each $N_M$ for a total of 420000 samples, half broken and half
unbroken. This is called $\mathrm{Cl_{NM}}$, and it is further divided
in two subsets, with $80\%$ of samples used for training the GNN
classifier and $20\%$ for standard testing on aggregate families (or
sizes) statistically homogeneous with those seen during the
training.\\ To test the GNN-classifier model on aggregates of families
never seen during the training, we consider a dataset containing
aggregates with different number of primary particles, namely $N_M =
256, 640$; this database, called $\mathrm{Cl_{heter}}$ and containing
120000 realizations, is also well balanced among the two aggregate
families, and in broken/unbroken samples. \\ Given the strong
dependence of the maximal tensile force on the aggregate size, the
training of the new GNN Classifier is done selecting random thresholds
for each aggregate family; $N_M$-specific thresholds are extracted
from the cumulative PDF of the maximal tensile force, and vary in the
range [0.3-0.9] of the CDF. \\ The results are summarised in Table
\ref{table:NM}: first we observe that the GNN-Classifier when tested
on aggregate families seen during the training (but clearly with
different velocity gradients and rotation realizations) behaves very
well with an average success probability of $87\%$; second, we observe
that the GNN-Classifier maintains such performances also when
operating on aggregates of sizes and maximal tensile force
distributions never seen during the training. Moreover, the success
probability and F$_1$-score are basically independent of $N_M$,
confirming the robustness of the performances. These results point to
the strong capability of learning and generalising of our model.

\section{Conclusion}
\label{sec:final}
We have investigated the ability of Graph Neural Networks to predict
the rupture of isostatic aggregates by hydrodynamical stresses in
three-dimensional turbulent suspensions. While we do not {\it
  interpret} our models \citep{Mehlig2021} - something that goes well
beyond the limits of our work-, we can assess their validity and
investigate their failures.\\ In particular, we have proposed two
distinct implementations of the GNNs, targeting classification or
regression tasks.  We showed that we generally obtain very good
accuracy in the breakup prediction, and that models tend to fail for
force values very close to the rupture thresholds. Clearly, the
threshold region is the one where making correct predictions is more
difficult. In applications where the goal is to avoid rupture, this
may not be a problem since the GNN prediction would represent an upper
bound to a conservative estimation; for applications targeting the
formation of small from large aggregates, GNN results should be
interpreted with additional care.\\ Concerning generalisation of model
behaviour, we showed that the implemented GNNs exhibit very good
accuracy when the test-set characteristics differ from those seen
during the training. This applies when considering random threshold
values for the rupture for the classifier model, and when considering
new geometrical shapes for the regression model. Similarly, very good
performances have been obtained when training the model on
heterogeneous in sizes aggregates, and testing it on aggregates of
sizes never seen during the training. \\ Our study opens the way to
further applications in the field of colloidal suspensions, or
microplastics modeling in complex flows. A promising avenue for future
studies is training a GNN model with full temporal histories of stress
and forces. By doing so, we could produce synthetic trajectories of
aggregate dynamics (see e.g. \cite{Li2024}), complemented by the
tensile forces, and give dynamical predictions, substituting numerical
integration in flows at different/higher Reynolds numbers.\\ Another
interesting direction is to develop more
  sophisticated GNN architectures, exploiting for instance attention
  mechanisms not only to predict the maximal tensile force value,
$F_{\text{max}}$, but also the bond where this is realized, or to
classify the rupture and identify the broken bond.  This would
eventually speed up the computation of breakup events and the
aggregate size distribution in numerical simulations with multiple
aggregates. A long-term goal targets the modeling of the interplay of
aggregation/fragmentation processes by data-driven methods. \\

\section{Appendix A}
\label{sec:appendix_grad}
The datasets listed in Table \ref{tab:1} have been prepared as follows.\\

{\bf Turbulent velocity gradients}: We perform Direct Numerical
Simulations of Homogeneous and Isotropic, forced three-dimensional
turbulence for the fluid velocity ${\bf v}({\bf x},t)$. Once the
Eulerian flow is statistically stationary, we integrate the Lagrangian
trajectories of tracer particles, $d{\bf X}/dt = {\bf v}({\bf
  X}(t),t)$. Along the tracer trajectories, we store all components of
the fluid velocity gradient tensor at the tracer position, $\partial_i
v_j (X(t))$. Then along each trajectory we select 2048 velocity
gradient realizations, sufficiently separated in time (about one
large-scale eddy-turn-over-time) such that they are uncorrelated. So
doing, we have built up a collection of independent and identically
distributed (with the correct turbulent statistics) velocity gradient
realizations. We remark that DNS are needed to obtain the intermittent
velocity gradient statistics that is observed in three-dimensional
turbulence \citep{frisch1995turbulence}. 

{\bf Aggregates}: We consider aggregates with the same fractal
dimension $D_F=1.9$ and number of primary particles $N_M=384$, but
different geometrical arrangements. We take each aggregate geometry
and rotate it by a random angle in the 3D space, and store independent
orientations of the same aggregate. At the end, we have a collection
of aggregates with independent orientations and independent
realizations of the turbulent velocity gradients. In this set-up, any
time is equivalent. For each of these, we compute the flow induced
forces/torques within the Stokesian Dynamics approach. These data are
stored and constitute the starting DB.

{\bf Breakup}: Having computed the tensile force distribution over all
samples with the Stokesian Dynamics, we identify rupture threshold
values. Then to each sample aggregate it is associated a rupture
threshold: since we know from SD the maximal tensile force for the
specific aggregate sample and we have a rupture value threshold, we
know if the aggregate wrt that threshold is a broken one (provided it
has one, and only one, internal bond with tensile force above the
threshold) or unbroken one (provided it has all internal bonds with
the tensile forces below threshold).

\section*{CRediT authorship contribution statement} 
Authors equally contributed to this work.

\section*{Declaration of competing interest}
The authors declare that they have no known competing financial interests or personal relationships that could have appeared to
influence the work reported in this paper.

\section*{Acknowledgements} M. C. and A. L. acknowledge support by the European Union Next Generation plan, through the Italian Ministry of University and Research (MUR) Piano Nazionale di Ripresa e Resilienza (PNRR), project title: 'CN00000013 – National Centre for HPC, Big Data and Quantum Computing'. M.B. acknowledge support by the Fondo Italiano per la Scienza 2022-2023 (FIS2) CUP E53C24003760001 A.L. is grateful to Gautier Verhille and Michael Wilczeck for useful discussions.\\

\section*{Data availability}
Data will be made available on request.
\bibliographystyle{elsarticle-harv} 
\bibliography{references}
\end{document}